# A carpet cloak for heat flux and temperature field

Tianzhi Yang[1]*,    Fei Chen[1],    Lujun Huang[2]

1. School of Aerospace Engineering, Shenyang Aerospace University, Shenyang 110136, China

2. National Laboratory for Infrared Physics, Shanghai Institute of Technical Physics, Shanghai, 200083, China

**Abstract**

Based on transformation optics (TO) theory, we present a new carpet device that can be used to thermally protect a region from the invasion of external heat flux. The designed device is termed "thermal carpet", which provides considerable cloaking effect. The cloaking performance for heat flux originating from different directions are analyzed. Unlike most thermal cloak designs reported in the literature, the material parameters are constant with position throughout the cloak, indicating that only one type of metamaterial composite to fabricate such a carpet is required.

The transformation optics (TO) has been proposed as a powerful tool to control over wave propagation. The concept was initially proposed in the context of electromagnetics[1,2], and later extended to acoustics [3-8]. One of the exciting applications of TO is "cloak", a coating shell can guide the propagation of light and acoustic waves. As a result, a region inside the shell become invisible.

Later, TO inspired many theoretical and experimental approaches for controlling dc magnetic field [9], elastic wave [10-13] and matter waves [14]. Very recently, the TO theory was also extended to thermodynamics[15-19]. Although heat flux is not a real wave because it does not transport energy as

---

[1]* Corresponding author

*Email addres*s: yangtz@me.com

1  normally waves do, the thermal cloak by using TO were still achieved. This is because the heat

2  conduction equation, also known as the heat diffusion equation, is invariance under coordinate

3  transformation[3]. These pioneer works open possibilities for cloaking and focusing heat flux. On

4  the other hand, we all know that polyurethane is often used for thermal insulation in various

5  engineering applications. However, its thermal conductivity is not low enough to totally shield

6  heat flux. Thus, the transformation thermodynamics may provide a new strategy to manipulate

7  heat flux.

8  However, such above mentioned heat flux cloaks generally require spatially varying

9  constitutive parameters, even results in extremely complex material parameters. Thus, such

10 devices have to be very inhomogeneous and anisotropic. Moreover, to the best of our knowledge,

11 the method of ground cloak has already been proved easier to realize in practice due to its property

12 that it can be created through quasiconformal mapping, and therefore the anisotropy can be

13 minimized inside the carpet [20]. The carpet cloak has be taken in order to demonstrate the cloak

14 concept for electromagnetic wave[21-24] and acoustics [25-26] but not for controlling heat flux. Thus in

15 this letter, we introduce the ground cloaking labeled "thermal carpet" or "thermal ground cloak"

16 able to manipulate heat flux.

17 The material parameters of the thermal carpet will be derived based on the TO theory. Thermal

18 conduction is the moment of a heat flux flows from a high temperature region toward a low

19 temperature region. Here we start from the thermal conduction equation without the source term

20 $$\nabla(-\kappa\nabla T)=0 \qquad (1)$$

21 where $\kappa$ is the thermal conductivity and $T$ is the temperature. In this study, the analysis is

22 restricted to a two dimensional case.

23 Figure 1 shows the detailed structure of the thermal carpet. The shaded region is the designed

24 carpet and the bottom triangular shape is the space needing to protect. For obtaining a ground

25 plane cloak, we consider the following transformations

26 for $-a \leq x_1 \leq 0$

27 $$x_1' = x_1 \qquad (2a)$$

1 $$x'_2 = \frac{\tau d}{a} x_1 + (1-\tau)x_2 + \tau d \tag{2b}$$

2 $$x'_3 = x_3 \tag{2c}$$

3 for $0 \leq x_1 \leq a$

4 $$x'_1 = x_1 \tag{3a}$$

5 $$x'_2 = -\frac{\tau d}{a} x_1 + (1-\tau)x_2 + \tau d \tag{3b}$$

6 $$x'_3 = x_3 \tag{3c}$$

7 where $a$, $d$ and $\tau$ are the geometric parameters, as shown in Figure 1. Note that equations (3) and

8 (4) are linear transformations. According to the TO theory, the effective thermal conductivity can

9 be derived by using the following equation

10 $$\tilde{\kappa}(x') = \frac{A\kappa(x)A^T}{\det(A)} \tag{4}$$

11 where $A$ is the Jacobian transformation matrix with elements defined by

12 $$A = \frac{\partial x'_i}{\partial x_i} \quad i=1, 2, 3 \tag{5}$$

13 By using the mapping functions (2)-(3), the heat conduction equation (1) is mapped into:

14 $$\nabla(-\tilde{\kappa}\nabla T)=0 \tag{6}$$

15 where the transformed thermal conductivity tensor $\tilde{\kappa}$ can be calculated by using equations (4) and

16 (5), the results are

17 for $-a \leq x_1 \leq 0$

18 $$\tilde{\kappa} = \frac{1}{\Delta}\begin{bmatrix} 1 & S \\ S & S^2 + \Delta^2 \end{bmatrix} \tag{7}$$

19 for $0 \leq x_1 \leq a$

20 $$\tilde{\kappa} = \frac{1}{\Delta}\begin{bmatrix} 1 & -S \\ -S & S^2 + \Delta^2 \end{bmatrix} \tag{8}$$

21 where $S = \tau d / a$, $\Delta = 1-\tau$ $\quad (1 \leq \tau \leq 1)$

22 Equations (7) and (8) are the expressions of the required physical effective material parameters of the thermal carpet. *More importantly, unlike most thermal cloak designs reported in the literature, the material parameters are constant with position throughout the cloak, indicating that only one type of metamaterial composite to fabricate such a thermal carpet is required.* Such carpet may realized by using materials composed of layers of isotropic material, solid inclusion or homogenization. Thus, comparing with previous work, this flat carpet is easier to realize physical cloak for heat flux.

The performance of the proposed cloak is quantitatively examined by using finite element method. As a example, we choose $a$=1m, $d$=3m and $\tau$=0.5. For comparison, we first consider the case without cloaking or the thermal carpet is made of isotropic material. The initial temperatures on the top and bottom of the external rectangular shape are chosen to be 100K and －100K. Due to the temperature difference within the rectangular region, heat flux will flow from top (hot) toward bottom (cold). This transport process takes place until the system has become uniform temperature throughout. In Figure 2, the white lines denote the isothermal lines. It is seen that the distribution of isothermal lines are uniform. With and without the isotropic device, temperature distribution is not alerted.

In contrast, we consider the coating carpet is made of metamaterial and set the same initial temperature difference. The thermal conductivity $\tilde{\kappa}$ of the carpet is spatially varied according to Equations (7) and (8). It is seen in Figure 3 that the thermal carpet greatly change the distribution of temperature field and the triangular region are thermally shielded. As a result, the thermal flux can be manipulated and the isothermal lines are bended in the carpet. Inside the cloaked region, the temperature at position $(x,y)$=(0, 0.5) is $T$=－86.2K which is much colder than external values. Thus, such device could be used to protect a certain region when a temperature gradient is suddenly imposed and minimized the temperature variation in the cloaked region.

On the other hand, the top and boundary boundaries are chosen to be －100K and 100K. Thus heat flux will flow from bottom (hot) to top (old). The stationary temperature profile inside the rectangular region is shown in Figure 4, it is seen that the triangular region is also cloaked and the temperature at position $(x,y)$=(0, 0.5) is $T$=86.9K. Outside this region, the temperature is much lower. This result suggests that the thermal carpet also can be used to keep warm in the cloaked region.

As a specially case, we examine the cloaking performance for heat flux transports between left and right sides. In Figure 5, the arrows are proportionally plotted according to the magnitude of heat flux. The solid purple lines denote the isothermal lines. It is seen that the isothermal lines are bended and most heat flux is transported. However, it is seen that there is a few thermal flux flows inside the cloaked region, indicating that the thermal carpet, for this case, is not perfect. This is because the thermal carpet has two points contacting the ground and they cannot totally shield the heat flux.

Moreover, the time dependent cloaking performance for $a$=1m, $d$=4m and $\tau$=0.5.is studied. Snapshots of heat distribution at $t$=1s is presented in Figure 6 (a), it is observed that the temperature inside the cloaked region at position (0, 0.5) is $T$=−33.3K. As time increases, as shown in Figure 6 (b), this temperature at position $(x,y)$=(0, 0.5) becomes $T$=−60.9K. As time increases, this cloaking region becomes colder and the temperature eventually stays at $T$=86.9K.

In conclusion, based on coordinate transformation method, a ground plane cloak for heat flux is presented. The coating carpet can keep warm/cold in the protected region for heat flux coming originating from different directions. For heat flux transports between left and right sides, the cloaking performance is not perfect and there is a small temperature variation inside the cloaked region. In general, such device could be used to protect a certain region from the invasion of the eternal heat flux. More importantly, comparing with previous work, this flat carpet is easier to realize physical thermal cloak. This is because the material parameters of our carpet are constant with position throughout the cloak, indicating that only one type of metamaterial composite to fabricate such thermal carpet is required.

## Acknowledgments

This work was supported by the National Natural Science Foundation of China (No. 11202140, 11172010 and 10702045).

1  **Figure Captions**

2  **Figure 1** Schematic of the ground plane cloak

3  **Figure 2** Temperature profile without cloaking with *a*=1m, *d*=3m and *τ*=0.5.

4  **Figure 3** Cloaking performance. The top and boundary boundaries are chosen to be 100K and

5  -100K.

6  **Figure 4** Cloaking performance. The top and boundary boundaries are chosen to be -100K and

7  100K.

8  **Figure 5** Cloaking performance between left and right boundaries (a) Heat flux transports from

9  left to right (b) Heat flux transports from right to left.

10  **Figure 6** The time dependent cloaking process.

11
12
13
14
15
16
17
18
19
20
21
22
23
24

1 **Figures**

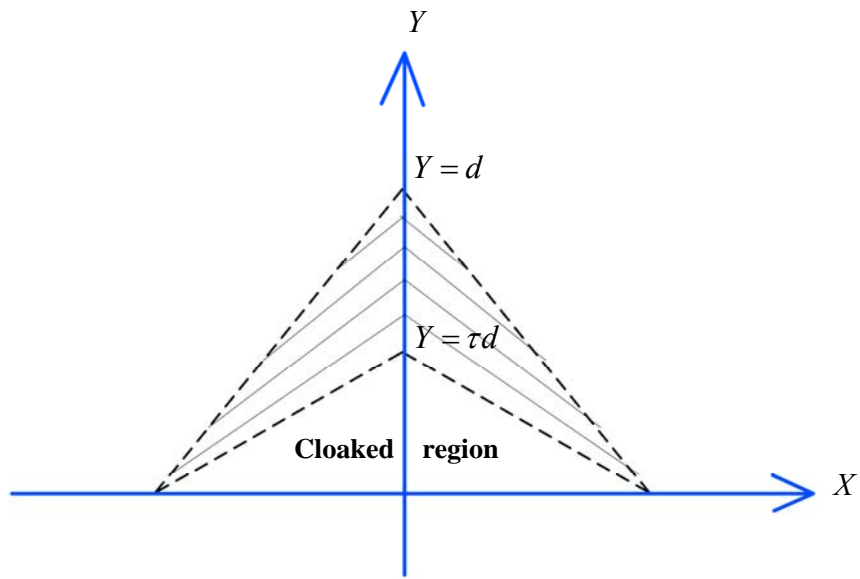

3
4 Figure 1

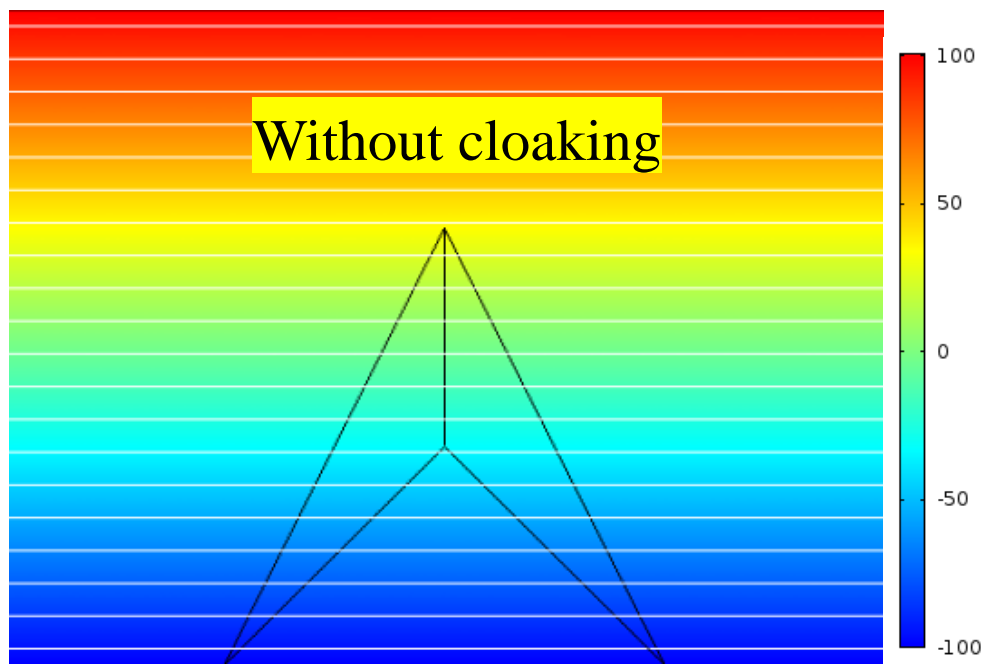

6
7 Figure 2

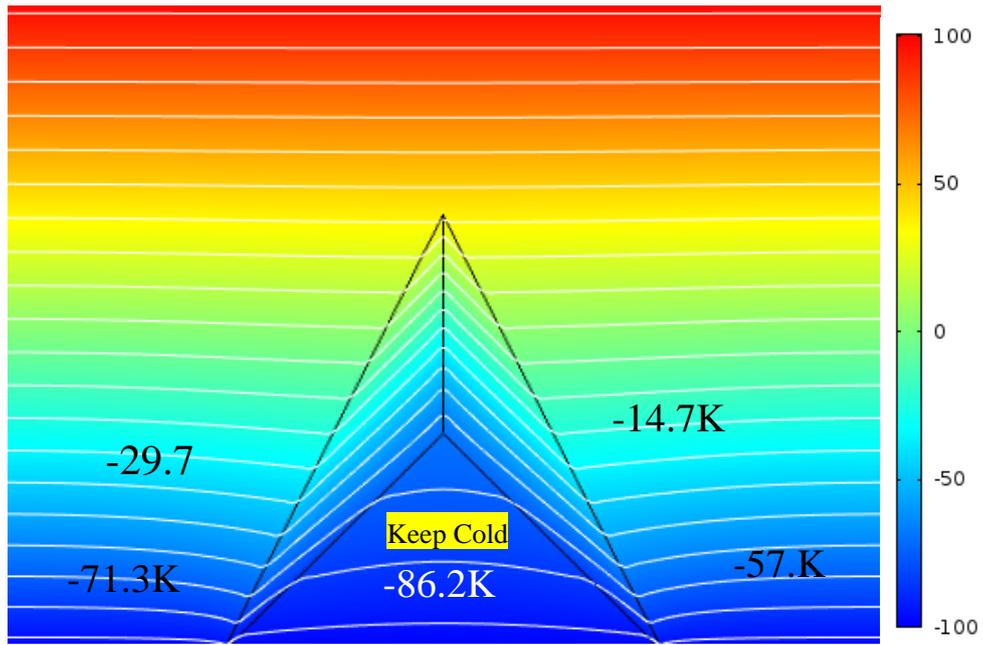

Figure 3

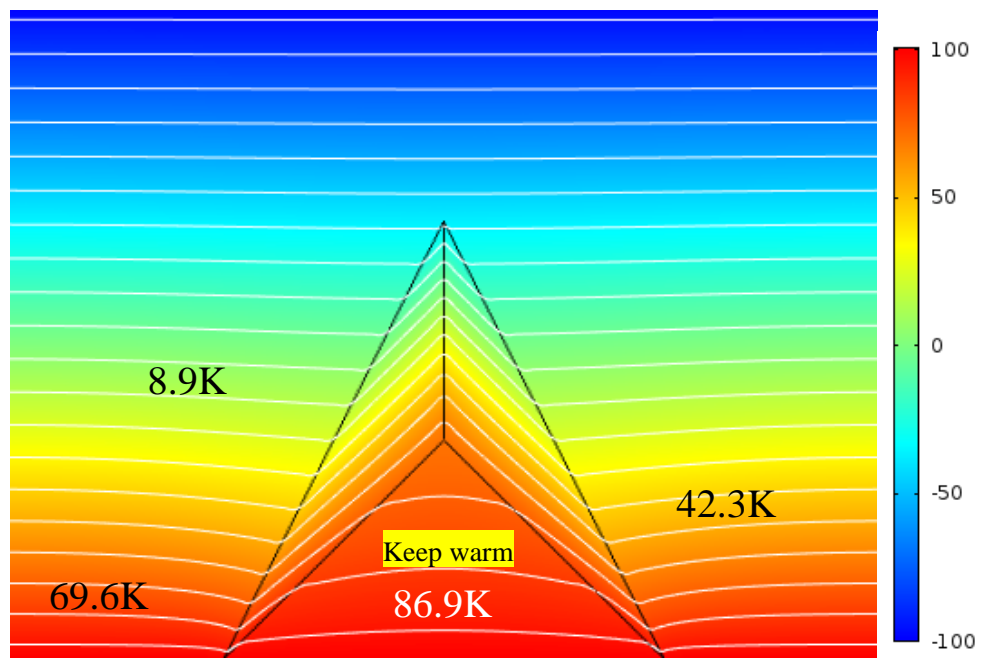

Figure 4

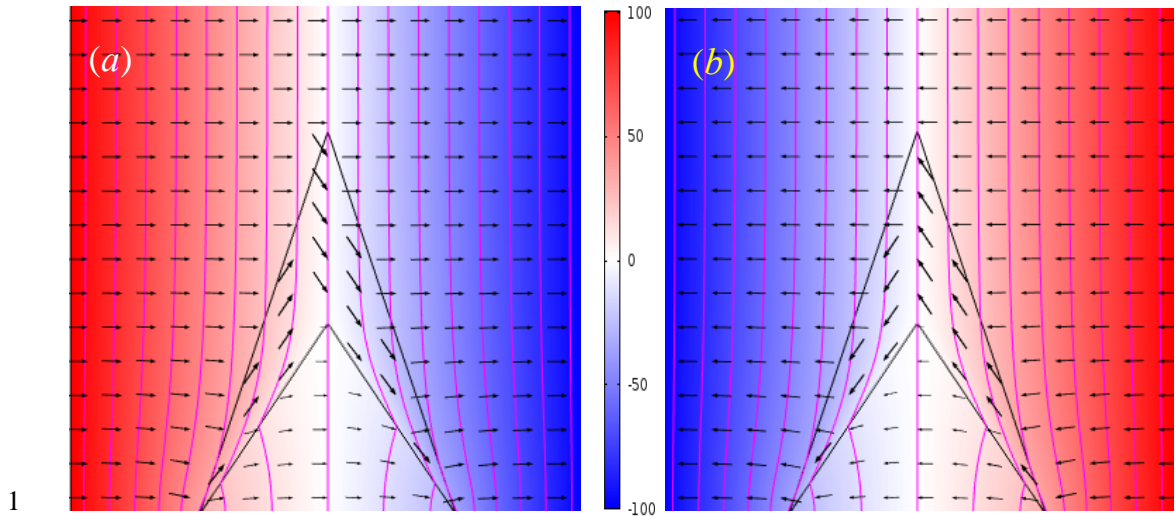

Figure 5

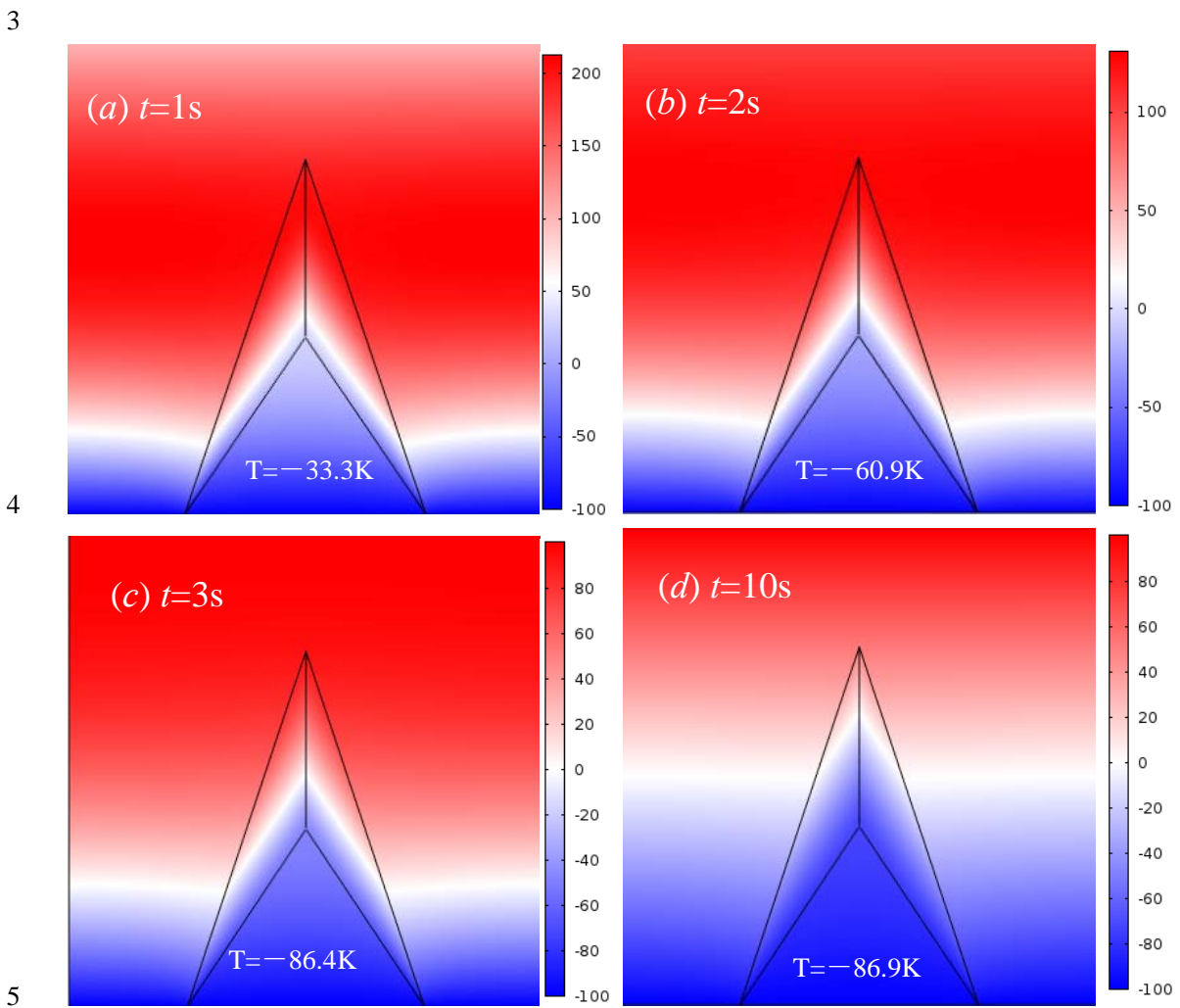

Figure 6